\documentclass{svproc}

\usepackage[utf8]{inputenc}
\usepackage{amsmath}
\usepackage{graphicx}
\usepackage[misc]{ifsym}
\usepackage{url}
\usepackage{hyperref}

\usepackage{xcolor}
\begin{document}
\mainmatter              
\title{The \texttt{wisdom\_of\_crowds}: an efficient, philosophically-validated, social epistemological network profiling toolkit}
\toctitle{The {wisdom\_of\_crowds}: an efficient, philosophically-validated, social epistemological network profiling toolkit}
\titlerunning{The \texttt{wisdom\_of\_crowds} social epistemological network profiling toolkit}  
%
\author{Colin Klein\inst{1} \and Marc Cheong\inst{2} \and Marinus Ferreira\inst{3} \and Emily Sullivan\inst{4} \and Mark Alfano\inst{3} }
\authorrunning{Colin Klein et al.} 
%
\tocauthor{Colin Klein, Marc Cheong, Marinus Ferreira, Emily Sullivan, and Mark Alfano}
\institute{The Australian National University, ACT, Australia\\
\email{colin.klein@anu.edu.au}
\and
University of Melbourne, VIC, Australia\\
\email{marc.cheong@unimelb.edu.au}
\and
Macquarie University, NSW, Australia
\and
Eindhoven University, Eindhoven, The Netherlands }

\maketitle              
\begin{abstract}
The epistemic position of an agent often depends on their position in a larger network of other agents who provide them with information. In general, agents are better off if they have diverse and independent sources. Sullivan et al. \cite{SullivanVulnerability20} developed a method for quantitatively characterizing the epistemic position of individuals in a network that takes into account both diversity and independence; and   presented a proof-of-concept, closed-source implementation on a small graph derived from Twitter data \cite{SullivanVulnerability20}. This paper reports on an open-source re-implementation of their algorithm in Python, optimized to be usable on much larger networks. In addition to the algorithm and package, we also show the ability to scale up our package to large synthetic social network graph profiling, and finally demonstrate its utility in analyzing real-world empirical evidence of `echo chambers' on online social media, as well as evidence of interdisciplinary diversity in an academic communications network.
\end{abstract}

{\footnotesize \color{blue}
\subsection*{Note About This Preprint}
This is a preprint of the following chapter: Colin Klein \& Marc Cheong \& Marinus Ferreira \& Emily Sullivan \& Mark Alfano,
"The wisdom\_of\_crowds: an efficient, philosophically-validated, social epistemological network profiling toolkit", 
published in "Complex Networks \& Their Applications XI: Proceedings of The Eleventh International Conference on Complex Networks and their Applications: COMPLEX NETWORKS 2022 - Volume 1", 
edited by Hocine Cherifi, Rosario Nunzio Mantegna, Luis M. Rocha, Chantal Cherifi, Salvatore Micciche, 2022, Springer reproduced 
with permission of Springer. 

The final authenticated version is available online at: \url{https://link.springer.com/chapter/10.1007/978-3-031-21127-0_6}.
}

\section{Introduction}
\label{s:Introduction}
Most of what we know we know because we learned about it from other people. \emph{Social epistemology} is the subfield of philosophy that studies how knowledge and justification depend on the testimony of others transmitted through social networks \cite{goldman1999knowledge}.  A focus on networks has been influential because it allows philosophers to connect their concerns to the substantial body of empirical and simulation work on real-world networks and their graph-theoretic properties. 

Sullivan et al. \cite{SullivanVulnerability20} presented a method for quantitatively characterizing the epistemic position of individuals in a network. Broadly speaking, individuals are in a better epistemic position if they are receiving information from \emph{diverse} and \emph{independent} sources, with the more diversity and independence the better. This method was based on, amongst others, the \textit{Wisdom of Crowds} hypothesis, that the aggregated judgements of many individuals can systematically be more accurate than the judgements of those individuals taken singly \cite{Surowiecki2005}. They then operationalized these two concepts in a way that allowed them to provide an interesting profile of a small 185-member Twitter community \cite{SullivanVulnerability20}. That work relied on a bespoke, closed-source codebase. As it was built as a proof-of-concept, it was also not optimized in ways that naturally scaled to larger networks. This  made it difficult to apply the technique to other datasets, such as networks from other social media sites, or networks created from artificial social simulation algorithms, e.g. Laputa \cite{olsson2011simulation}: all of which are of interest to both philosophers and computer scientists alike. That work's codebase had an emphasis on the generation of Java-based visualizations---using a combination of several platforms and toolchains--- which does not lend itself to convenient large-scale network analysis, due to performance limitations.

To make this tool more widely available to researchers, we therefore present \verb+wisdom_of_crowds+, a complete ground-up re-implementation in Python of the core Sullivan et al. \cite{SullivanVulnerability20} concepts. The code is optimized to deal with larger networks. It also includes some standardized helper functions to allow for coordinating results between research groups and data scientists. We have made the code for this package open source, under the \textit{GNU General Public License 3.0}\footnote{Full license terms can be found at \url{https://www.gnu.org/licenses/gpl-3.0.en.html}.}, on GitHub (\url{https://github.com/cvklein/wisdom-of-crowds/}). It has also been accepted to the Python Package Index (PyPI, at \url{http://pypi.org/project/wisdom-of-crowds}), and is available for any user to install via \verb+pip+\footnote{It can be installed by the Python community via: \texttt{pip install wisdom\_of\_crowds}}.

As much as possible, we have relied on open-source Python packages such as \verb+networkx+ \cite{schult2008exploring} and \verb+matplotlib+ \cite{Hunter2007} as they have been rigorously tested and are freely available for auditing and peer review. For good practices in verification and validation, the \verb+pytest+ \cite{pytestxy}  library is used to provide a unit-testing framework. 
Having created this new implementation, we sought out to deploy it in our investigation of contemporary social network data, to provide a data-driven perspective to complement existing conceptual and theoretical work. 

This paper thus aims to present two key findings:
\begin{itemize}
    \item the core concepts of the \texttt{wisdom\_of\_crowds} package, including compatibility with (and optimization for) contemporary network-related datasets and packages in Python. This includes improvements to the base \cite{SullivanVulnerability20} algorithm, by clearly defining the bounds (and justifications for) parameters used, as well as suggested extensions including the $h$-measure derived from \cite{Hirsch2005}.
    \item application of \texttt{wisdom\_of\_crowds} on simulated large networks; to investigate its feasibility/performance
    \item application of \texttt{wisdom\_of\_crowds} on actual real-world networks---Twitter discourse about the Black Lives Matter (BLM) movement between January and July 2020 centering around the May 25th 2020 murder of George Floyd \cite{Alfano2022BLM}; and the \textit{email-Eu-core} network on communications patterns in ``a large European research institution'' \cite{eucore,Yin2017Core,Leskovec2007Core}---to corroborate theoretical network epistemological findings on modern-day social networks centered on current phenomena.
\end{itemize}

\section{Background and Methods}
\label{s:BackgroundAndMethods}

\subsection{Core concepts}
\label{ss:CoreConcepts}

\textit{Epistemology} is a branch of philosophy which, simply put, ``is concerned with how people should go about the business of trying to determine what is true'' \cite{SEP2019}.
As per our Introduction (Section \ref{s:Introduction}), \emph{social epistemology} concerns the testimony of others embedded in social contexts \cite{goldman1999knowledge}; in contrast with `individual' epistemology which concerns how an individual conducts reasoning, abstracted away from their ``social environment''  \cite{SEP2019}.

In recent years, social epistemologists have moved away from considering dyadic relationships between individuals to consider the ways in which social epistemic \emph{networks} shape the information we receive \cite{o2019misinformation,alfano2020humility}. Consider an epistemic network $G$ where nodes are epistemic agents and edges represent the relationship of receiving information via testimony. `\textit{Testimony}' is used broadly in social epistemology for any way in which one source delivers information to another, and includes speech, writing, and other forms of media. All things being equal, a node is better off receiving information from more and more diverse nodes.  However, testimony is often transmitted in chains, and this transmission need carry only the content of the information, not (meta-)information about the original source or the intermediate links. This complicates the position of any individual who is trying to learn from multiple sources.  For example, a piece of gossip heard from two people seems more reliable than from one, but that reliability is undermined if both heard it from the same person  \cite{AlfanoGossip17}.

\subsection{Sullivan et al. (2020)'s operationalizations}
\subsubsection{Defining the $m,k$ observer:}
Following \cite{SullivanVulnerability20}, we say that a node $n$ is an $m,k$-observer just in case it receives information from a set of at least $k$ different nodes which are pairwise at least $m$ steps away from one another, when considered on the subgraph of $G$ that does not contain $n$. If $G$ is directed, then candidate sources must be at least $m$ steps away in both directions. The removal of $n$ from consideration in the case of distances is necessary for directed graphs, as otherwise all sources to $n$ are trivially at most 2 steps apart; we carry over that requirement to undirected graphs as well. 

In this work, we assume $1\leq m\leq 5$, as most real life networks have length $6$ paths between most arbitrarily chosen nodes \cite{MilgramThe-small67}. We bound $2 \leq k \leq 5$, because a node with a single source is in a very poor epistemic position with respect to diversity of input. Note that it is a consequence of the definition that if $n$ is an $m,k$-observer, it is also both an $m-1,k$-observer and an $m,k-1$-observer (assuming $m-1$ and $k-1$, respectively, are permissible values). 

Given this definition, the core concepts in \cite{SullivanVulnerability20} are defined as follows. 

\subsubsection{$S(n)$, independence of sources.}
$S(n)$ gives a measure of the independence of sources to node $n$.  Consider the set $s$ of possible pairs $(m,k)$ for which $n$ is an $m,k$-observer. Then define

\begin{equation}
    S(n)= \begin{cases}
    0 & \text{if $s=\emptyset$}\\
    \max\{mk : (m,k)\in s\} & \text{otherwise}
    \end{cases}
\end{equation}
In other words, $S(n)$ is just the largest $mk$ such that the $n$ is an $m,k$-observer. If $S$ has $0$ or $1$ nodes as sources, they are considered as being in an epistemically bad position, and so $S(n)==0$. Note that given this definition, possible $S$ values do not increase smoothly. Given the bounds set out above, $S(n)\in \{0,2,3,4,5,6,8,9,10,12,15,16,20,25\}$.

\subsubsection{$D(n)$, diversity of sources.}
$D(n)$ measures the diversity of the sources that contribute information to $n$. Let each node $i$ be associated with a set $a_i$ of epistemically-relevant attributes. These might be group affiliations, topics of interest, scientific approaches, political leanings, and so on. Let $s$ be the set of $n$'s sources. Then define

\begin{equation}
    D(n) = \mid \bigcup \{a_i : i\in s\}\mid
\end{equation}
That is, $D$ gives the number of distinct \emph{types} of information that feed into $n$. 

\subsubsection{$\pi(n)$, epistemic position.}
Finally, as the epistemic position of a node is a function of both the diversity and independence of sources, we define 

\begin{equation}
    \pi (n) = S(n)D(n)
\end{equation}

\subsection{Caveats and Considerations}
\label{ss:CaveatsAndConsiderations}

There are a few notes to make about the implications of the Sullivan et al. \cite{SullivanVulnerability20}  core concepts in real social networks.

Regarding $m,k$-observers, while higher rankings are better and all the nodes with a specific value of $m$ and $k$ are members of an equivalence class, the framework does not posit which of two $m,k$-observers is better positioned if one has a higher $m$ value and the other a higher $k$ value---for instance, whether $2,3$-observers are better- or worse-placed than $3,2$-observers. The framework thus does not provide a total order but instead provides a collection of \textit{partial} orders. 

Intuitively, for instance, a $1,2$-observer is worse placed than a $2,3$-observer despite having neither an $m$ or a $k$ value in common, because a $2,3$-observer is better placed than a $2,2$-observer when considering $k$ values, which in turn is better placed than a $1,2$-observer when comparing $m$ values.
Thus, a node's $S$ value does provide a way to more easily compare the epistemic position of nodes, by combining their $m$ and $k$ values into a single value. There are concerns with it, though. 

The choice of multiplying $m$ and $k$ values together to come to a single measure is ultimately arbitrary. The $S$ value contains less information than the $m$ and $k$ values does, because it fails to preserve the difference between being a $2,3$-observer and being a $3,2$-observer (the former has fewer independent sources, but these have a greater degree of independence from each other than in the latter's). Simultaneously, it posits that a node which is a $3,3$-observer is determinately better placed than either a $4,2$-observer or a $2,4$-observer, but worse than a $5,2$-observer or a $2,5$-observer. It is unclear why we should believe that this is true in general. 

In short, while a node's $S$ value is one measure of the independence of its  sources, it is unclear why we should use this measure rather than another. This was the impetus for us introducing a further measure, the \verb+h_measure(n)+ \cite{Hirsch2005} which returns the highest $h$ such that $n$ is a $h,h$-observer, as discussed below.

\subsection{Our Re-Implementation}
\label{ss:OurRe-Implementation}

The core of the \verb+wisdom_of_crowds+ is a class \verb+Crowd+\footnote{Due to space constraints, we only discuss our substantial contributions within this paper. The full documentation for all implemented methods is available at \url{https://github.com/cvklein/wisdom-of-crowds/blob/main/docs/wisdom\_of\_crowds.py.md}}. \verb+Crowd+ is initialized with a \textit{NetworkX} graph (encapsulating the social network's edges and nodes), and provides various functions to calculate the metrics defined above. 

Calculating whether a node is an $m,k$-observer combines multiple shortest-path problems with clique-finding problems. Na\"ive approches to the latter have complexity $O(n^k)$ \cite{VassilevskaEfficient09}. Given that we are considering unweighted paths, the shortest-path problem has a reasonably efficient linear-time solution, but the  requirement to remove $n$ from the distance calculations also means that network shortest paths cannot simply be calculated at the outset. In the worst case scenario, they  must be recalculated for each pair of sources for each node.  

Hence this is a computationally difficult problem to brute-force. In practice, efficient caching and testing of seen paths plus greedy $k$-clique algorithms means that worst-case performance can often be avoided for realistic networks: we \textit{memoize} (i.e., cache) intermediate path values, trading-off space for a reduced processing time. As the envisioned use case for large graphs is for one-shot batch processing, our code allows for easy multithreading or multiprocessing (e.g. using \textit{multiprocessing} or \textit{concurrent.futures}) allowing it to attain substantial performance gains, in conjunction with the memoization mechanism. 

The $m,k$-observer functionality is the basis for calculating $D$, $S$, and $\pi$. $D$ is calculated on node attributes, and users can specify the appropriate key for the attribute. If a single attribute is supplied, $D$ is calculated using the singleton set containing that attribute. 

In addition to the standard measures, we also introduce an improvement: the \verb+h_measure()+ of a node $n$ as the smallest $x$ such that $n$ is an $x,x$-observer; comparable to the standard definition of Hirsch's \textit{h}-index in citation practices \cite{Hirsch2005}. As suggested by \cite{SullivanVulnerability20}, being a 3,3-observer is the minimal secure epistemic position, and the use of a single non-multiplicative standard may be useful for some cases.  

Finally, the package includes two helper functions to allow for comparable reporting and display across different users. \verb+iteratively_prune_graph()+ takes a \textit{NetworkX} graph, removes small-degree nodes, small-weight edges, and takes the largest connected component in what remains, iterating this process until the graph is stable. The thresholds are parameterized; the default is for $indegree+outdegree \leq 1$ and no edge culling, as per \cite{SullivanVulnerability20}. In the spirit of scientific reproducibility, \verb+make_sullivanplot()+ makes a summary figure of a whole network in the style of \cite[refer to their Figure 7]{SullivanVulnerability20}. It can produce standalone plots or return a subplot in a specified \verb+matplotlib+ axis. 

\section{Experimental Results}
\label{s:Results}

\subsection{Efficiency Tests}
\label{ss:EfficiencyTests}
Firstly, to benchmark the ability of \verb+wisdom_of_crowds+ on different magnitudes of nodes, typical of modern-day datasets, we sought to batch calculate $S$ for different magnitudes of node sizes ($|N|$), given various probabilities of edge connections.

\begin{figure}[ht!]
\centering
\includegraphics[width=\textwidth]{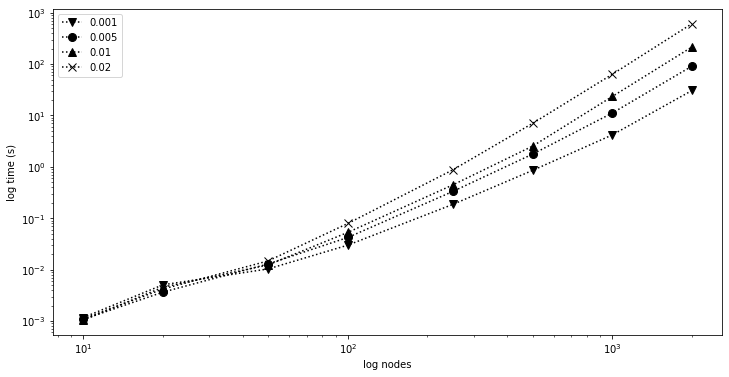}
\caption{Timing curve for random graphs by number of nodes. Different markers represent different connection probabilities for nodes.}
\label{fig:timing}
\end{figure}

Figure \ref{fig:timing} shows the efficiency of batch calculating $S$ for each node of a \verb+Crowd+ on random graphs with varying parameters for probability of edge connection. 
Random directed graphs were generated using the \verb+networkx+ generator\\ \verb+fast_gnp_random_graph()+ with the parameters indicated in the figure. Timing was done using the python $\verb+timeit+$ package over a single iteration.

As the log-linear plot in Figure \ref{fig:timing} shows, there is a roughly exponential relationship between the number of nodes and runtime, with the exponent a function of the number of edges. This suggests that the efficiency of our code approaches what would be expected given the fundamental complexity of the clique-finding problem. Note that the exponential growth means that the boundary between computationally tractable and intractable graphs can be relatively tight. Judicious pruning often makes a difference. 

Further, note that this was achieved in standard Python operating conditions, i.e. in the absence of any multiprocessing/multithreading support (see also Section \ref{ss:OurRe-Implementation}). Performance gains will be attained for batch calculations on high-CPU (16-or-more CPU cores) systems.

\subsection{Application on Social Network Data: BLM on Twitter}
\label{ss:ApplicationBLMOnTwitter}
Secondly, as part of our investigation of real-world phenomena, in line with \cite{SullivanVulnerability20}, we replicate their findings on a real-world Twitter retweet network to examine the information-sharing dynamics during the Black Lives Matter movement.
An earlier version of this dataset was used by \cite{SullivanVulnerability20} who were able to examine a network of 185 nodes. For further information on the dataset\footnote{We queried the Twitter Streaming API with a series of Black Lives Matter (BLM)-related keywords, hashtags, and short expressions in a window between January and July 2020.The dataset comprised $\sim$4.6M original tweets between January 13th and July 18th and $\sim$94.5M retweets from January 18th to July 23rd; these tweets were produced by $\sim$2.0M distinct authors.} used, see \cite{Alfano2022BLM}. For brevity, the key details are summarised here.

A \textit{retweet network} \cite{SullivanVulnerability20,Alfano2022BLM} was generated: i.e., a weighted directed network where nodes are authors and the weight of an edge from node $u$ to node $v$ represents the number of times that user $v$ retweeted user $u$. 
We took the largest connected component of this graph as the starting point for cluster-analysis \cite{leiden,igraph}.
Following \cite{Alfano2022BLM}, We found first-level clusters using Modularity Vertex Partitioning, preserving clusters with more than 10\% of the original nodes. This gave 4 clusters, covering 83\% of the graph. Next, we manually inspected the 100 most-influential nodes within each group, characterizing the communities as Activists, Center-Left Democrats, Republicans, and a set of ``Boosters'' who mainly amplified the content of the first two groups. Topic models were fit using \textit{scikit-learn}'s non-negative matrix factorization (NMF), fit on a \textit{tf-idf} representation of the Twitter text (post-sanitization) with \verb+min_df=0.05+.
Finally, we used \verb+iteratively_prune_graph()+ with a node and weight threshold of 3,  which resulted in a tractable subgraph with $|N|=16249$ nodes and $|E|=145246$ edges. This subgraph had very little representation from the `booster' group, so they were omitted from further analysis. By comparison with \cite{SullivanVulnerability20}, our analysis is an increase in the number of nodes by a magnitude of $\sim$100$\times$. Batch processing took about 6.25 hours on a 2017 desktop iMac. 

\begin{figure}[!htb]
\centering
\includegraphics[width=\textwidth]{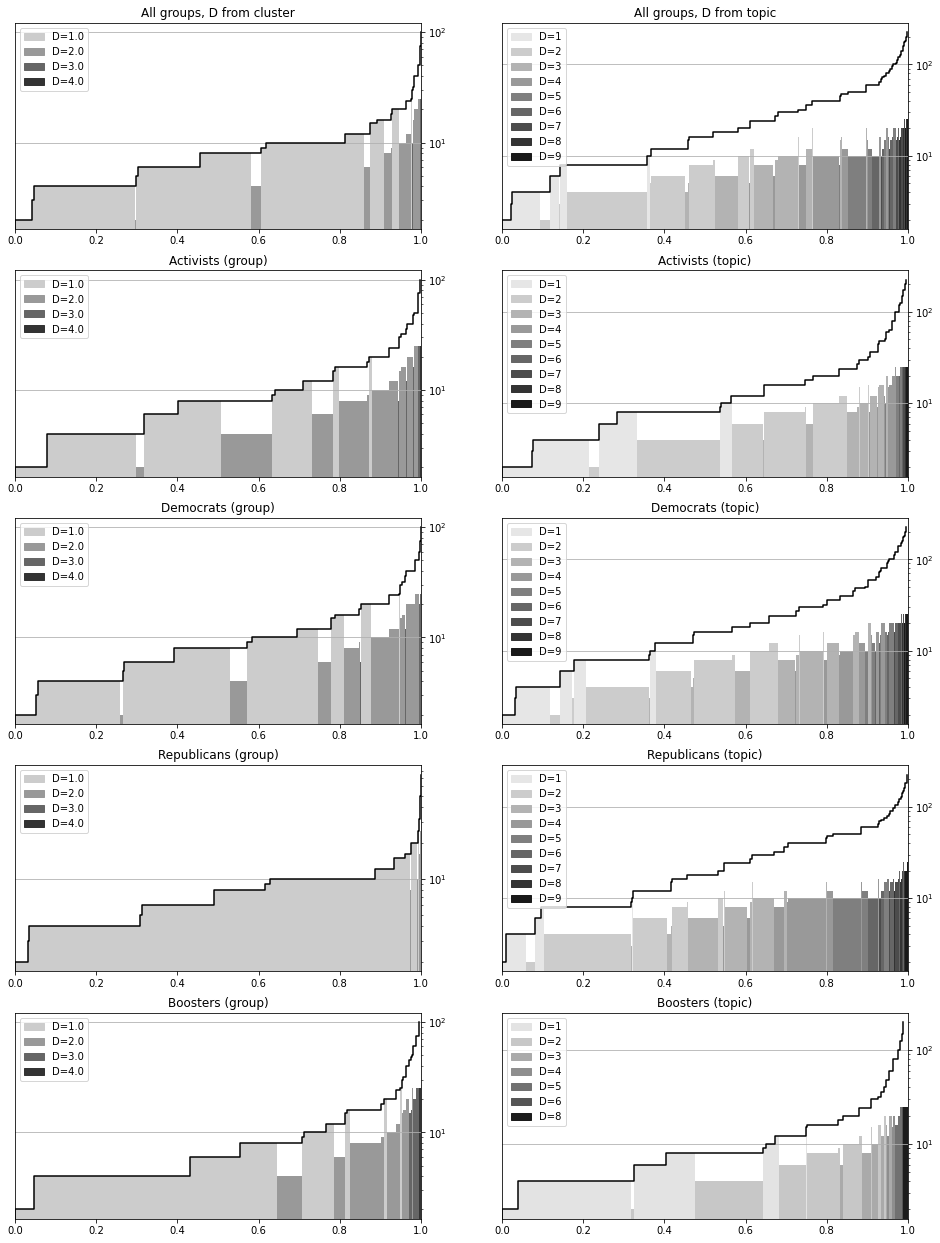}
\caption{Profile plots for entire network and subgroups looking at clusters (left) and topics (right). X axis is proportion of total, Y axis shows both S (height of bars) and $\pi$ (black line), plotted on a log scale.}
\label{fig:twitter}
\end{figure}

Figure \ref{fig:twitter} plots $S$, $D$, and $\pi$ for the current experiment. For the left side of \ref{fig:twitter}, $D$ was calculated using the cluster identity of the node. For the right side, $D$ was calculated using the \verb+argmax+ of the fitted and normalized $W$ matrix for the topic model. This gives the topic that is most distinctive of each user's tweets.  We examine both the network as a whole and three identified subclusters in the graph. The left half of Figure \ref{fig:twitter} shows  $S$, $D$, and $\pi$ for the network, where $D$ is calculated via the subcluster identity of sources. The right half of Figure \ref{fig:twitter} recalculates $D$ and $\pi$ based on a 9-topic NMF topic model of aggregated tweets (compared with the 3-topic model of \cite{SullivanVulnerability20}).

The results illustrate the utility of profiling networks using our toolkit. On the left, one can see that Republicans appear to be in the worst epistemic position in terms of the other subgroups with which they interact: they have a generally low $D$, suggesting that they tend to listen mostly to in-group members. However, they have a relatively high $S$ and therefore a $\pi$ comparable to other subgroups. Compare this to the topic-wise graph, in which Republicans have a a relatively high diversity for \emph{topics}, one at least as good as other groups. The activist group shows something of the inverse pattern. That is, they show a more varied range of $S$ and $D$ values for group-group interactions, but a comparatively lighter graph with fewer topics for the broad span. 

These results might suggest that both groups are part of `echo chambers' \cite{AlfanoTechnological18,nguyen2020echo}, but in different ways: the right tends to be a monoculture socially but a polyculture topically, with a converse pattern on the left. Finally, we note that all subgroups, in both domains of measurement, tend to have an $S<10$ for more than half the population. This replicates the observations of \cite{SullivanVulnerability20}, in which most participants end up in a comparatively poor epistemic position. However, most groups tend to contain at least a small sub-population which is well-connected and often with a relatively high $D$. We note that this is especially the case with our `Booster' group, a small subset who seemed to be mainly concerned with amplifying and relaying  the content of other groups.

\subsection{Application on Communication Network Data: \textit{email-Eu-core}}
\label{ss:ApplicationEuCore}
Finally, we apply our \verb+wisdom_of_crowds+ analysis onto the \textit{email-Eu-core} network, an existing dataset curated by \cite{Leskovec2007Core,Yin2017Core} on the Stanford Large Network Dataset Collection \cite{eucore}. The rationale behind this is to apply our social epistemological lens to analyze an existing network which is hitherto (to our knowledge) never been examined using the tools we have at hand.

Briefly, \textit{email-Eu-core} consists of ``email data from a large European research institution'' \cite{eucore}, represented as a digraph where an edge ($u$, $v$) exists ``if [researcher] $u$ sent [researcher] $v$ at least one email''. The beauty of   \textit{email-Eu-core} lies in the fact that it only considers internal communications in the institution, ignoring any noise resulting from possible non-academic communication originating from/in reply to outside actors; and that the ground-truth membership of each node has already been established, i.e., as belonging to any ``one of 42 departments at the research institute'' \cite{eucore}.

\begin{figure}[!htb]
\centering
\includegraphics[width=1.1\textwidth]{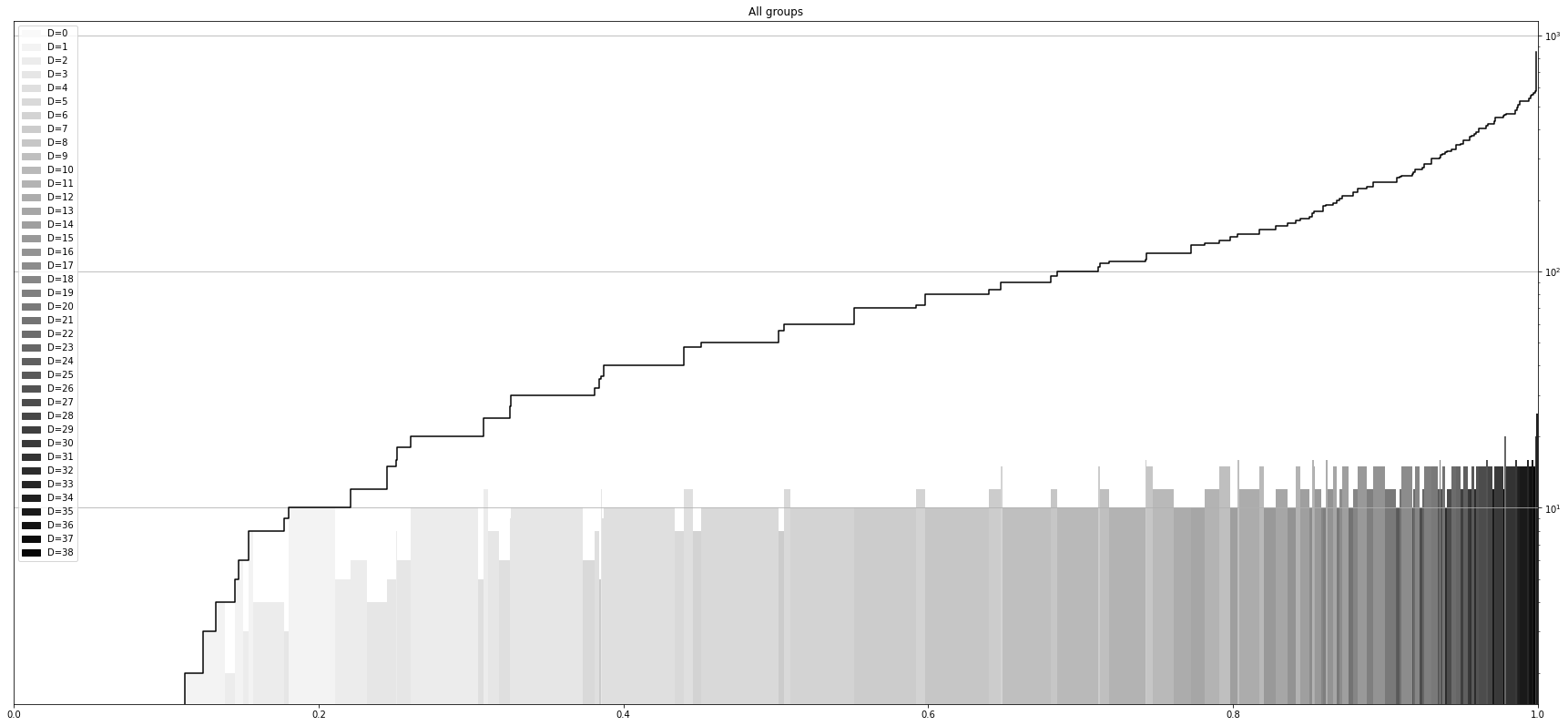}
\caption{Profile plots for the entire \textit{email-Eu-core} network \cite{Leskovec2007Core,Yin2017Core,eucore}. The X axis is proportion of total, Y axis shows both S (height of bars) and $\pi$ (black line), plotted on a log scale.}
\label{fig:eucore}
\end{figure}

Compared to more modern social networks, \textit{email-Eu-core} has a comparatively small $|N|=1005$ and $|E|=25571$. However, again, we note that this is already $\sim$10$\times$ the magnitude as compared to the hitherto state-of-the-art \cite{SullivanVulnerability20}. The total running time on such a dataset is comparatively trivial (less than 5 minutes). As we do not have e-mail text associated with the original nodes in this dataset, we sought only to analyze the `overall epistemic picture' of the network, as a whole. In the same vein as Section \ref{ss:ApplicationBLMOnTwitter}, Figure \ref{fig:eucore} illustrates the profile plot for \textit{email-Eu-core}.

Again, the utility of our approach is evident. Given the results, we observe that the distribution of $S$ is fairly consistent for all the nodes (researchers) in the research institution's academic network. However, the progressively darker bars illustrate that researchers who have connections with a more diverse amount of departments (thereby maximizing $D$) can vastly optimize their epistemic position. Roughly, about $\sim40\%$ of researchers -- i.e., the right-most data points in Figure \ref{fig:eucore} -- have a $\pi$ of about 100 or more, despite having roughly the \textit{same} $S$ values.

To our knowledge, this is the first time empirical social epistemological analyses in the spirit of \cite{SullivanVulnerability20} have been conducted on such email networks.

\section{Discussion}

 Our results show that it is possible to replicate the methodology used by \cite{SullivanVulnerability20} to larger networks, and that insights about the relative epistemic positions of different communities within a network can be drawn from plotting these parameters. As our package and its dependencies are all open source, this makes it possible for researchers in a range of fields (including philosophy, psychology, sociology, anthropology, communications,  and network science) both to conduct new research and to reanalyze networks that they have previously studied. 

We anticipate that future research will expand the types of social networks under study. Other sources from online social media such as Facebook, Reddit, and YouTube all seem to be viable candidates for study.  Considering offline epistemic networks would be especially valuable, such as those found in the landmark Bernard-Kilworth-Sailer (\textit{BKS}) analyses of social networks \cite{Bernard1979}, as their structure may be interestingly different from the structures found online. \verb+wisdom_of_crowds+ also allows us to conduct research on epistemic network simulations, created with tools such as Laputa \cite{olsson2011simulation}, which can quickly scale up to thousands of nodes. We expect that studies of friend networks, organizational networks in industry and the military, networks of sources used by journalists, criminal cartel networks, and academic citation networks would prove valuable. 

Moving beyond that, it would be interesting to study networks with more than one type of testimonial edge (e.g., public communications versus private ones). One intriguing hypothesis is that these may differ in structure even if they contain the same nodes, and that individuals who are central in public networks but peripheral in private networks (or vice versa) would tend to play unique roles in the social epistemology of those networks. For instance, someone who is privately in communication with a very large number of others but not publicly visible is in a position to exert influence because the others may assume that they have a much better epistemic position than they actually do.

The exploratory profiling made possible by our tool reveals patterns of epistemic isolation and interaction across real-world networks, and suggests possibilities for more specific analyses.  By providing it to the community at large, we hope to facilitate further modeling of epistemic networks across a variety of domains. 

\section*{Funding}
Work on this paper was supported by ARC Grant DP190101507 (to C.K.\ and M.A.) and by Templeton Grant 61378 (to M.A.).

This research was supported by use of the Nectar Research Cloud and by Melbourne Research Cloud (University of Melbourne) (to M.C.).  The Nectar Research Cloud is a collaborative Australian research platform supported by the NCRIS-funded Australian Research Data Commons (ARDC).

\end{document}